\documentclass[10pt]{scrartcl}
\usepackage{hyperref}
\usepackage{epsfig}
\usepackage{times}
\usepackage[numbers]{natbib}
\usepackage{amsmath}
\usepackage{amsbsy}
\usepackage{amsfonts}
\usepackage{multicol}
\usepackage[latin1]{inputenc}
\usepackage{url}

\typearea{12}
 \providecommand{\email}[1]{E-mail:
\href{mailto:#1}{\texttt{#1}}}
\providecommand{\dprod}{\! \cdot \!}%
\providecommand{\wprod}{\! \wedge \!}
\begin{document}
%

\title{The null subspace of $\mathcal{G}_{4,1}$ as source of the main physical
theories}
\author{\textbf{J. B. Almeida}\\ \normalsize
{Universidade do Minho, Departamento de F\'isica,}\\\normalsize {4710-057
Braga, Portugal.}\\\normalsize \email{bda@fisica.uminho.pt}}

\pagestyle{myheadings} \markright{The null subspace of $\mathcal{G}_{4,1}$ as
source of the main physical theories \hfill J. B. Almeida }


\date{\normalsize }

%
%

%
\maketitle

\begin{abstract}                
The relationship between geometry and physics is probably stronger in General
Relativity (GR) than in any other physics field. It is the author's belief that
a perfect theory will eventually be formulated, where geometry and physics
become indistinguishable, so that the complete understanding of space
properties, together with proper assignments between geometric and physical
entities, will provide all necessary predictions.

We don't have such perfect theory yet, however the author intends to
show that GR and Quantum Mechanics (QM) can be seen as originating
from properties of the null subspace of 5-dimensional space with
signature $(-++++)$, together with its associated geometric algebra
$\mathcal{G}_{4,1}$. The space so defined is really 4-dimensional
because the null condition effectively reduces the dimensionality by
one. Besides generating GR and QM, the same space generates also
4-dimensional Euclidean space where dynamics can be formulated and
is quite often equivalent to the relativistic counterpart. Euclidean
relativistic dynamics resembles Fermat's principle extended to 4
dimensions and is thus designated as 4-Dimensional Optics (4DO).

In this presentation the author starts with the geometric algebra
$\mathcal{G}_{4,1}$ with imposition of the null displacement length condition
and derives the method to transpose between the metrics of GR and 4DO; this
transition is proven viable for stationary metrics. It is hopeless to apply
Einstein type equations in 4DO, for the simple reason that a null Ricci tensor
always leads to a metric diverging to infinity. The author uses geometric
arguments to establish alternative equations which are solved for the case of a
stationary mass and produce a solution equivalent to Schwarzschild's metric in
terms of PPN parameters.

As a further development, the author analyses the case of a monogenic function
in $\mathcal{G}_{4,1}$. The monogenic condition produces an equation that can
be conveniently converted into Dirac's, with the added advantage that it has
built in standard model gauge group symmetry.
\end{abstract}
%

\begin{multicols}{2}
\section{Introduction}
According to general consensus any physics theory is based on a set of
principles from which predictions are derived using established mathematical
derivations; the validity of such theory depends on agreement between
predictions and observed physical reality. In that sense this paper does not
formulate physical theories because it does not presume any physical
principles; for instance it does not assume speed of light constancy or
equivalence between frame acceleration and gravity. This is a paper about
geometry; all along the paper, in several occasions, a parallel is made with
the physical world by assigning a physical meaning to geometric entities and
this allows predictions to be made. However the validity of derivations and
overall consistency of the exposition is independent of prediction correctness.

The only postulates in this paper are of a geometrical nature and
can be summarised in the definition of the space we are going to
work with; this is the 4-dimensional null subspace of the
5-dimensional space with signature $(-++++)$. The choice of this
geometric space does not imply any assumption for physical space up
to the point where geometric entities like coordinates and geodesics
start being assigned to physical quantities like distances and
trajectories. Some of those assignments will be made very soon in
the exposition and will be kept consistently until the end in order
to allow the reader some assessment of the proposed geometric model
as a tool for the prediction of physical phenomena. Mapping between
geometry and physics is facilitated if one chooses to work always
with non-dimensional quantities; this is done with a suitable choice
for standards of the fundamental units. From this point onwards all
problems of dimensional homogeneity are avoided through the use of
normalising factors listed below for all units, defined with
recourse to the fundamental constants: $\hbar \rightarrow$ Planck
constant divided by $2 \pi$, $G \rightarrow$ gravitational constant,
$c \rightarrow$ speed of light and $e \rightarrow$ proton charge.

\begin{center}
\begin{tabular}{c|c|c|c}
Length & Time & Mass & Charge \\
\hline & & & \\

$\displaystyle \sqrt{\frac{G \hbar}{c^3}} $ & $\displaystyle \sqrt{\frac{G
\hbar}{c^5}} $  & $\displaystyle \sqrt{\frac{ \hbar c }{G}} $  & $e$
\end{tabular}
\end{center}

This normalisation defines a system of \emph{non-dimensional units} (Planck
units) with important consequences, namely: 1) All the fundamental constants,
$\hbar$, $G$, $c$, $e$, become unity; 2) a particle's Compton frequency,
defined by $\nu = mc^2/\hbar$, becomes equal to the particle's mass; 3) the
frequent term ${GM}/({c^2 r})$ is simplified to ${M}/{r}$.

4-dimensional space can have amazing structure, providing countless
parallels to the physical world; this paper is just a limited
introductory look at such structure and parallels. The exposition
makes full use of an extraordinary and little known mathematical
tool called geometric algebra (GA), a.k.a.\ Clifford algebra, which
received an important thrust with the works of David Hestenes
\cite{Hestenes84}. A good introduction to GA can be found in
\citet{Gull93} and the following paragraphs use basically the
notation and conventions therein. A complete course on physical
applications of GA can be downloaded from the internet
\cite{Lasenby99}; the same authors published a more comprehensive
version in book form \cite{Doran03}. An accessible presentation of
mechanics in GA formalism is provided by \citet{Hestenes03}.

\section{Introduction to geometric algebra}
We will use Greek characters for the indices that span 1 to 4 and Latin
characters for those that exclude the 4 value; in rare cases we will have to
use indices spanning 0 to 3 and these will be denoted with Greek characters
with an over bar. Einstein's summation convention will be adopted as well as
the compact notation for partial derivatives $\partial_\mu = \partial/\partial
x^\mu$. The geometric algebra of the hyperbolic 5-dimensional space we want to
consider $\mathcal{G}_{4,1}$ is generated by the frame of orthonormal vectors
$\{\mathrm{i},\sigma_\mu \}$, $\mu = 1 \ldots 4$, verifying the relations
\begin{eqnarray}
    && \mathrm{i}^2  = -1,\\ && \mathrm{i} \sigma_\mu + \sigma_\mu \mathrm{i}
    =0,\\
    && \sigma_\mu
    \sigma_\nu + \sigma_\nu \sigma_\mu  = 2 \delta_{\mu \nu}.
\end{eqnarray}
We will simplify the notation for basis vector products using multiple indices,
i.e.\ $\sigma_\mu \sigma_\nu \equiv \sigma_{\mu\nu}$. The algebra is
32-dimensional and is spanned by the basis
\begin{itemize}
\item 1 scalar, { $1$}
\item 5 vectors, { $\{\mathrm{i},\sigma_\mu \}$}
\item 10 bivectors (area), { $\{\mathrm{i} \sigma_\mu,\sigma_{\mu\nu} \}$}
\item 10 trivectors (volume), { $\{\mathrm{i} \sigma_{\mu\nu},
    \sigma_{\mu\nu\lambda}  \}$}
\item 5 tetravectors (4-volume), { $\{\mathrm{i}I, \sigma_\mu I \}$}
\item 1 pseudoscalar (5-volume), { $I \equiv \mathrm{i}\sigma_1 \sigma_2
    \sigma_3\sigma_4$}
\end{itemize}
Several elements of this basis square to unity:
\begin{equation}
    (\sigma_\mu)^2 =  (\mathrm{i} \sigma_\mu)^2=
    (\mathrm{i}\sigma_{\mu\nu})^2 =(\mathrm{i}I)^2 =1;
\end{equation}
and the remaining square to $-1$:
\begin{equation}
    \mathrm{i}^2 = (\sigma_{\mu\nu})^2 = (\sigma_{\mu\nu\lambda})^2 =
    (\sigma_\mu I)^2 = I^2=-1.
\end{equation}
Note that the symbol $\mathrm{i}$ is used here to represent a vector
with norm $-1$ and must not be confused with the scalar imaginary,
which we don't usually need. Note also that the pseudoscalar $I$
commutes with all the other basis elements while being a square root
of $-1$ and plays the role of the scalar imaginary in complex
algebra.

The geometric product of any two vectors $a = a^0 \mathrm{i} + a^\mu
\sigma_\mu$ and $b = b^0 \mathrm{i} + b^\nu \sigma_\nu$ can be decomposed into
a symmetric part, a scalar called the inner product, and an anti-symmetric
part, a bivector called the exterior product.
\begin{equation}
    ab = a \dprod b + a \wprod b,~~~~ ba = a \dprod b - a \wprod b.
\end{equation}
Reversing the definition one can write interior and exterior products as
\begin{equation}
    a \dprod b = \frac{1}{2}\, (ab + ba),~~~~ a \wprod b = \frac{1}{2}\, (ab -
    ba).
\end{equation}

When a vector is operated with a multivector the inner product reduces the
grade of each element by one unit and the outer product increases the grade by
one. There are two exceptions; when operated with a scalar the inner product
does not produce grade $-1$ but grade $1$ instead, and the outer product with a
pseudoscalar is disallowed.

\section{Displacement and velocity}
Any displacement in this 5-dimensional hyperbolic space can be defined by the
displacement vector
\begin{equation}
    \label{eq:displacement}
    \mathrm{d}x =\mathrm{i} \mathrm{d}x^0 + \sigma_\mu \mathrm{d}x^\mu;
\end{equation}
and the null space condition implies that $\mathrm{d}x$ has zero length
\begin{equation}
    \mathrm{d}x^2 = \mathrm{d}x \dprod \mathrm{d}x = 0;
\end{equation}
which is easily seen equivalent to either of the relations
\begin{equation}
\begin{split}
    \label{eq:twospaces}
    (\mathrm{d}x^0)^2 &= \sum (\mathrm{d}x^\mu)^2;\\
    (\mathrm{d}x^4)^2 &= (\mathrm{d}x^0)^2 - \sum (\mathrm{d}x^j)^2.
\end{split}
\end{equation}
These equations define the metrics of two alternative 4-dimensional spaces, one
Euclidean the other one Minkowskian, both derived from the null 5-dimensional
subspace.

A path on null space does not have any affine parameter but we can use Eqs.\
(\ref{eq:twospaces}) to express 4 coordinates in terms of the fifth one. We
will assign the letter $t$ and physical time to coordinate $x^0$ while the
letter $\tau$ and physical proper time are assigned to coordinate $x^4$; total
derivatives with respect to $t$ will be denoted by an over dot while total
derivatives with respect to $\tau$ will be denoted by a "check", as in
$\check{f}$. Dividing both members of Eq.\ (\ref{eq:displacement}) by
$\mathrm{d}t$ we get
\begin{equation}
    \label{eq:euclvelocity}
    \dot{x} = \mathrm{i} + \sigma_\mu \dot{x}^\mu = \mathrm{i} + v.
\end{equation}
This is the definition for the velocity vector $v$; it is important
to stress again that the velocity vector defined here is a
geometrical entity and its possible relation to physical velocity is
a direct result of the coordinate assignments made above; if later
we were to find that the velocity vector bears no relation to
physical velocity only the assignments would have to be reviewed but
the mathematical deductions would retain their validity. The
velocity has unit norm because $\dot{x}^2 =0$; evaluation of
$v\dprod v$ yields the relation
\begin{equation}
    \label{eq:vsquare}
    v \dprod v = \sum (\dot{x}^\mu)^2 = 1.
\end{equation}
The velocity vector can be obtained by a suitable rotation of any of the
$\sigma_\mu$ frame vectors, in particular it can always be expressed as a
rotation of the $\sigma_4$ vector; we will make use of this possibility later
on.

At this point we are going to make a small detour for the first
parallel with physics. In the previous equation we replace $x^0$ by
the Greek letter $\tau$ and rewrite with $\dot{\tau}^2$ in the first
member
\begin{equation}
    \label{eq:dtau2}
    \dot{\tau}^2 = 1 - \sum (\dot{x}^j)^2.
\end{equation}
The relation above is well known in special relativity, see for
instance \citet{Martin88}; see also \citet{Almeida02:2} and
\citet{Montanus01} for parallels between special relativity and its
Euclidean space counterpart.\footnote{Montanus first proposed the
Euclidean alternative to relativity in 1991, nine years before the
author started independent work along the same lines.} We note that
the operation performed between Eqs.\ (\ref{eq:vsquare}) and
(\ref{eq:dtau2}) is a perfectly legitimate algebraic operation since
all the elements involved are scalars. Obviously we could also
divide both members of Eq.\ (\ref{eq:displacement}) by
$\mathrm{d}\tau$
\begin{equation}
    \check{x} =  \mathrm{i}\check{x}^0 + \sigma_j \check{x}^j + \sigma_4.
\end{equation}
Squaring the second member and noting that it must be null we obtain
$(\check{x}^0)^2 - \sum (\check{x}^j)^2 = 1$. This means that we can
relate the vector $\mathrm{i}\check{x}^0 + \sigma_j \check{x}^j$ to
relativistic 4-velocity, although the norm of this vector is
symmetric to what is usual in SR. The relativistic 4-velocity is
more conveniently assigned to the 5D bivector
$\sigma_4\mathrm{i}\check{x}^0 + \sigma_{4j} \check{x}^j$, which has
the necessary properties. The method we have used to make the
transition between 4D Euclidean space and Minkowski spacetime
involved the transformation of a 5D vector into scalar plus bivector
through product with $\sigma_4$; this method will later be extended
to curved spaces.

We will now define a new vector {$\mathrm{d}s$} related to displacement by the
scale factor {$n$}
\begin{equation}
    \label{eq:refindex}
    \mathrm{d}{s} = \mathrm{i}\mathrm{d}t + n \sigma_\mu \mathrm{d}{x}^\mu.
\end{equation}
In this way we are including the 4-dimensional analogue of a refractive index;
the previous equation is a generalisation of the 3-dimensional definition of
refractive index for an optical medium, which relates the optical path of light
in that medium to the geometric path. The factor $n$ used here scales the 4D
displacement vector $\sigma_\mu \mathrm{d}{x}^\mu$ and so it deserves the
designation of 4-dimensional refractive index; from now on we will drop the
"4-dimensional" qualification because the confusion with the 3-dimensional case
can always be resolved easily. The material presented in this paper is, in many
respects, a logical generalisation of optics to 4-dimensional space; so, even
if the paper is only about geometry, we will frequently use the designation
4-dimensional optics (4DO) when dealing with Euclidean 4-space.

Further generalisation of Eq.\ (\ref{eq:euclvelocity}) makes use of a tensor,
similar to the non-isotropic refractive index of optical media
\begin{equation}
    \label{eq:vgeneral}
    \mathrm{d}{s} = \mathrm{i}\mathrm{d}t + {n^\mu}_\nu \sigma_\mu
    \mathrm{d}{x}^\nu.
\end{equation}
The velocity is accordingly defined by $v = {n^\mu}_\nu \dot{x}^\nu
\sigma_\mu$. The same expression can be used with any orthonormal
frame, including for instance spherical coordinates, but for the
moment we will restrict our attention to those cases where the frame
does not rotate in a displacement in order to avoid having to derive
frame vectors when taking derivatives. This restriction poses no
limitation on the problems to be addressed but it is obviously
inconvenient when symmetries are involved and shall later be
relaxed.

The velocity can be given the more familiar form $v = g_\nu
\dot{x}^\nu$ if we define the \emph{refractive index frame}
\begin{equation}
    \label{eq:gmu}
    g_\nu = {n^\mu}_\nu \sigma_\mu.
\end{equation}
Obviously Eq.\ (\ref{eq:vgeneral}) implies that the velocity is
still a unitary vector and we can express this fact with through the
internal product with itself
\begin{equation}
    \label{eq:vsquaregen}
    v \dprod v = {n^\alpha}_\mu \dot{x}^\mu {n^\beta}_\nu \dot{x}^\nu
    \delta_{\alpha \beta}=1.
\end{equation}
Using Eq.\ (\ref{eq:vgeneral}) to evaluate $\mathrm{d}s^2 = 0$,
considering the definition (\ref{eq:gmu}) and denoting $g_{\mu \nu}
= g_\mu \dprod g_\nu$
\begin{equation}
    \label{eq:4dometric}
    (\mathrm{d} t)^2 = g_{\mu \nu} \mathrm{d}{x}^\mu \mathrm{d}{x}^\nu.
\end{equation}
This equation defines the metric of 4D space with signature $(++++)$, where $t$
is the geodesic arc length; this will be designated as 4DO metric because it
applies to 4-dimensional optics space.

In a similar way to what allowed us to derive 4DO and Minkowski spaces from the
null subspace condition, we will now show that general relativity (GR) metric
can also be derived from the same condition when the refractive index is
considered. In order to do this we define the reciprocal frame
$\{-\mathrm{i},g^\mu\}$ such that
\begin{equation}
    \label{eq:recframe}
    g^\mu \dprod g_\nu = {\delta^\mu}_\nu.
\end{equation}
Following the procedure outlined in \cite{Doran03} to determine the
reciprocal frame vectors we define the 4-volume tetravector
\begin{equation}
    V = \bigwedge_{\mu } g_\mu  = |V| \sigma_{1234},
\end{equation}
where the big wedge symbol is used to make the exterior product of
the $g_\mu$. The reciprocal frame vectors can then be found using
the formula \cite{Doran03}
\begin{equation}
    g^\nu = (-1)^\nu \bigwedge_{\nu \neq \mu} g_\mu V^{-1}.
\end{equation}

Use the reciprocal frame to multiply both members of Eq.\ (\ref{eq:vgeneral})
first on the right then on the left by $g^4$, simultaneously replacing $x^4$ by
$\tau$
\begin{align}
    \mathrm{d}s g^4 &= \mathrm{i}g^4 \mathrm{d}t + g_j g^4
    \mathrm{d}x^j + g_4 g^4  \mathrm{d}\tau;\\
    g^4 \mathrm{d}s &=
    g^4 \mathrm{i} \mathrm{d}t + g^4 g_j
    \mathrm{d}x^\mu + g^4 g_4  \mathrm{d}\tau.
\end{align}
Performing the inner product between the two equations and setting
$\mathrm{d}s^2$ to zero we get
\begin{equation}
    \label{eq:grmetric}
    (\mathrm{d}\tau)^2 = \frac{1}{g_{44}} (\mathrm{d}t)^2 - \frac{g_{j
    k}}{g_{44}}\mathrm{d}x^j    \mathrm{d}x^k ;
\end{equation}
which is recognizably a GR metric. Equation (\ref{eq:vgeneral}) can
then generate both 4DO and GR metrics, provided some conditions are
met; naturally the $g_\mu$ must be independent of $t$ if Eq.\
(\ref{eq:4dometric}) is to be taken as 4DO metric definition and
conversely they must not depend on $\tau$ if Eq.\
(\ref{eq:grmetric}) defines a GR metric. However we can say that for
static metrics at least we can convert between GR and 4DO.

\section{The sources of space curvature}
Equations (\ref{eq:4dometric}) and (\ref{eq:grmetric}) define two
alternative 4-dimensional spaces 4DO and GR respectively; in the
former $t$ is an affine parameter while in the latter it is $\tau$
that takes such role. Provided the metric is static the geodesics of
one space can be mapped one to one with those of the other and we
can choose to work on the space that best suits us.

The procedure to write the geodesic equations is the same in any curved space;
if we choose to work in 4DO this involves consideration of the Lagrangian
\begin{equation}
    L = \frac{g_{\mu \nu} \dot{x}^\mu \dot{x}^\nu}{2} = \frac{1}{2}\ .
\end{equation}
The justification for this choice of Lagrangian can be found in several
reference books but see for instance \citet{Martin88}. From the Lagrangian one
defines immediately the conjugate momenta
\begin{equation}
    v_\mu = \frac{\partial L}{\partial \dot{x}^\mu} = g_{\mu \nu} \dot{x}^\nu.
\end{equation}
Notice the use of the lower index ($v_\mu$) to represent momenta while velocity
components have an upper index ($v^\mu$). The conjugate momenta are the
components of the conjugate momentum vector $v = g^\mu v_\mu$ and from Eq.\
(\ref{eq:recframe})
\begin{equation}
    \label{eq:momentvel}
    v = g^\mu v_\mu = g^\mu g_{\mu \nu} \dot{x}^\nu = g_\nu \dot{x}^\nu.
\end{equation}
The conjugate momentum and velocity are the same but their
components are referred to the reciprocal and refractive index
frames, respectively.

The geodesic equations can now be written in the form of Euler-Lagrange
equations
\begin{equation}
    \dot{v}_\mu = \partial_\mu L;
\end{equation}
these equations define those paths that minimise $t$ when displacements are
made with velocity given by Eq.\ (\ref{eq:vgeneral}). Considering the parallel
already made with general relativity we can safely say that geodesics of 4DO
spaces have a one to one correspondence to those of GR in the majority of
situations.

We are going to need geometric calculus which was  introduced by
\citet{Hestenes84} as said earlier; another good reference is provided by
\citet{Doran03}. The existence of such references allows us to introduce the
vector derivative without further explanation; the reader should search the
cited books for full justification of the definition we give below. We define
two vector derivatives: the first one is represented by the symbol $\Box$ and
is referred to the reciprocal frame $g^\mu$ while the second one uses the
symbol $\nabla$ and is referred to the Euclidean frame $\sigma^\mu =
\sigma_\mu$
\begin{align}
    \Box &= g^\mu \partial_\mu; \\
    \nabla &= \sigma^\mu \partial_\mu;
\end{align}
the two vector derivatives are obviously related since $g^\mu$ can be expressed
in terms of $\sigma^\mu$.

The vector derivatives are vectors and as such they can be operated
with any multivector using the established rules; in particular the
geometric product of $\Box$ with a multivector can be decomposed
into inner and outer products. When applied to vector $a$ the result
is $(\Box\, a = \Box \dprod a + \Box \wprod a)$; the inner product
term is the divergence of vector $a$ and the outer product term is
the exterior derivative, related to the curl but contrary to the
latter it is usable in spaces of arbitrary dimension and is
expressed as a bivector. We also define curved and Euclidean
Laplacian as a result of multiplying each vector derivative with
itself; the result is necessarily a scalar
\begin{equation}
    \Box^2 = \Box \dprod \Box, \quad \nabla^2 = \nabla \dprod \nabla.
\end{equation}

Velocity is a vector with very special significance in 4DO space because it is
the unitary vector tangent to a geodesic; we therefore attribute high
significance to velocity derivatives, since they express the characteristics of
the particular space we are considering. When the Laplacian is applied to the
velocity vector this corresponds to the product of a scalar and a vector and
the result is necessarily a vector
\begin{equation}
    \label{eq:current}
    \Box^2  v = T.
\end{equation}
Vector $T$ is called the \emph{sources vector} and can be expanded into sixteen
terms as
\begin{equation}
    T = (\Box^2 {n^\mu}_\nu)
    \sigma_\mu \dot{x}^\nu = {T^\mu}_\nu \sigma_\mu \dot{x}^\nu .
\end{equation}
The tensor ${T^\mu}_\nu$ contains the coefficients of the sources vector and we
call it the \emph{sources tensor}; it is very similar to the stress tensor of
GR, although its relation to geometry is different. The sources tensor
influences the shape of geodesics as we shall see in one particularly important
situation.

Before we begin searching solutions for Eq.\ (\ref{eq:current}) we
note that it can be decomposed into a set of equations similar to
Maxwell's. Consider first the velocity derivative $\Box\, v = \Box
\cdot v + \Box \wedge v$; the result is a multivector with scalar
and bivector part $G = \Box\, v$. Now derive again: $\Box\, G = \Box
\cdot G + \Box \wedge G$; we know that the exterior derivative of
$G$ vanishes and the divergence equals the sources vector. Maxwell's
equations can be written in a similar form, as was shown in
\citet{Almeida04:2}; here the velocity was replaced by the vector
potential and multivector $G$ was replaced by the Faraday bivector
$F$; \citet{Doran03} offer similar formulation for spacetime.

Let us now concentrate on isotropic space, characterised by orthogonal
refractive index vectors $g_\mu$ whose norm can change with coordinates but is
the same for all vectors. Normally we relax this condition by accepting that
the three $g_j$ must have equal norm but $g_4$ can be different. The reason for
this relaxed isotropy is found in the parallel usually made with physics by
assigning dimensions $1$ to $3$ to physical space. Isotropy in a physical sense
need only be concerned with these dimensions and ignores what happens with
dimension 4. We will therefore characterise an isotropic space by the
refractive index frame $g_j = n_r \sigma_j$, $g_4 = n_4 \sigma_4$. Indeed we
could also accept a non-orthogonal $g_4$ within the relaxed isotropy concept
but we will not do so in this work.

We will only investigate spherically symmetric solutions independent of $x^4$;
this means that the refractive index can be expressed as functions of $r$ if we
adopt spherical coordinates. The vector derivative in spherical coordinates is
of course
\begin{equation}
\begin{split}
    \Box =& \frac{1}{n_r}\, \left(\sigma_r \partial_r + \frac{1}{r}\,
    \sigma_\theta \partial_\theta + \frac{1}{r \sin \theta}\, \sigma_\varphi
    \partial_\varphi \right) \\
    &   + \frac{1}{n_4}\, \sigma_4 \partial_4.
\end{split}
\end{equation}
The Laplacian is the inner product of $\Box$ with itself but the
frame vectors' derivatives must be considered; all the derivatives
with respect to $r$ are zero and the others are
\begin{alignat}{2}
    \partial_\theta \sigma_r &= \sigma_\theta,
     ~~ &\partial_\varphi \sigma_r &= \sin \theta \sigma_\varphi, \nonumber \\
    \partial_\theta \sigma_\theta &=
    -\sigma_r, ~~ &\partial_\varphi \sigma_\theta &= \cos \theta \sigma_\varphi, \\
    \partial_\theta \sigma_\varphi &= 0, ~~
    &\partial_\varphi \sigma_\varphi &= -\sin \theta\, \sigma_r - \cos \theta\, \sigma_\theta.
    \nonumber
\end{alignat}
After evaluation the curved Laplacian becomes
\begin{equation}
\begin{split}
    \label{eq:laplacradial}
    \Box^2 &= \frac{1}{(n_r)^2}\, \left(\partial_{rr} + \frac{2}{r}\, \partial_r -
    \frac{n'_r}{n_r}\, \partial_r  + \frac{1}{r^2}\, \partial_{\theta \theta}
     \right . \\
    & \left .
    +\frac{\cot \theta}{r^2}\, \partial_\theta
    + \frac{\csc^2 \theta}{r^2}\, \partial_{\varphi \varphi} \right) +
    \frac{1}{(n_4)^2}\, \partial_{\tau \tau}.
\end{split}
\end{equation}

The search for solutions of Eq.\ (\ref{eq:current}) must necessarily start with
vanishing second member, a zero sources situation, which one would implicitly
assign to vacuum; this is a wrong assumption as we will show. Zeroing the
second member implies that the Laplacian of both $n_r$ and $n_4$ must be zero;
considering that they are functions of $r$ we get the following equation for
$n_r$
\begin{equation}
    n^{''}_r + \frac{2 n'_r}{r} - \frac{(n'_r)^2}{n_r} = 0,
\end{equation}
with general solution $n_r = b \exp(a/r)$. It is legitimate to make $b =1$
because the refractive index must be unity at infinity. Using this solution in
Eq.\ (\ref{eq:laplacradial}) the Laplacian becomes
\begin{equation}
    \Box^2 = \mathrm{e}^{-a/r}\left(\mathrm{d}^2_r + \frac{2}{r}\,
    \mathrm{d}_r
     + \frac{a }{r^2}\, \mathrm{d}_r\right).
\end{equation}
When applied to $n_4$ and equated to zero we obtain solutions which impose $n_4
= n_r $ and so the space must be truly isotropic and not relaxed isotropic as
we had allowed. The solution we have found for the refractive index components
in isotropic space can correctly model Newton dynamics, which led the author to
adhere to it for some time \cite{Almeida01:4}. However if inserted into Eq.\
(\ref{eq:grmetric}) this solution produces a GR metric which is verifiably in
disagreement with observations; consequently it has purely geometric
significance.

The inadequacy of the isotropic solution found above for
relativistic predictions deserves some thought, so that we can
search for solutions guided by the results that are expected to have
physical significance. In the physical world we are never in a
situation of zero sources because the shape of space or the
existence of a refractive index must always be tested with a test
particle. A test particle is an abstraction corresponding to a point
mass considered so small as to have no influence on the shape of
space; in reality a point particle is a black hole in GR, although
this fact is always overlooked. A test particle must be seen as
source of refractive index itself and its influence on the shape of
space should not be neglected in any circumstances. If this is the
case the solutions for vanishing sources vector may have only
geometric meaning, with no connection to physical reality.

The question is then how should we include the test particle in Eq.\
(\ref{eq:current}) in order to find physically meaningful solutions.
Here we will make an \emph{ad hoc} proposal, without further
justification, because the author has not yet completed the work
that will provide such justification in geometric terms. The second
member of Eq.\ (\ref{eq:current}) will not be zero and we will
impose a sources vector based on the Euclidean Laplacian
\begin{equation}
    \label{eq:statpart}
    J = -\nabla^2 n_4 \sigma_4.
\end{equation}
Equation (\ref{eq:current}) becomes
\begin{equation}
    \label{eq:gravitation}
    \Box^2 v = -\nabla^2 n_4 \sigma_4;
\end{equation}
as a result the equation for $n_r$ remains unchanged but the equation for $n_4$
becomes
\begin{equation}
    n^{''}_4 + \frac{2 n'_4}{r} - \frac{n'_r n'_4}{n_r}
    = - n^{''}_4 + \frac{2 n'_4}{r}.
\end{equation}

When $n_r$ is given the exponential form found above, the solution
is $n_4 = \sqrt{n_r}$. This can now be entered into Eq.\
(\ref{eq:grmetric}) and the coefficients can be expanded in series
and compared to Schwarzschild's for the determination of parameter
$a$. The final solution, for a stationary mass $M$ is
\begin{equation}
    \label{eq:refind}
    n_r = \mathrm{e}^{2M/r},~~~~n_4 = \mathrm{e}^{M/r}.
\end{equation}

Equation (\ref{eq:gravitation}) can be interpreted in physical terms
as containing the essence of gravitation. When solved for
spherically symmetric solutions, as we have done, the first member
provides the definition of a stationary gravitational mass as the
factor $M$ appearing in the exponent and the second member defines
inertial mass as $\nabla^2 n_4$. Gravitational mass is defined with
recourse to some particle which undergoes its influence and is
animated with velocity $v$ and inertial mass cannot be defined
without some field $n_4$ acting upon it. Complete investigation of
the sources tensor elements and their relation to physical
quantities is not yet done; it is believed that the 16 terms of this
tensor have strong links with homologous elements of stress tensor
in GR but this will have to be verified.

\section{Wave optics in 4D}
In the previous paragraphs we have seen how relativistic dynamics
can be derived from an extension of geometric optics into 4D
Euclidean space and the question naturally arises if a similar
extension of wave optics can provide any new insight into physics;
this section will give us some idea of the possibilities opened by
such approach.

Any 4D wave must verify the general wave equation
\begin{equation}
    \label{eq:4dwave}
    \Box^2 \psi = \frac{\partial^2 \psi}{\partial t^2}.
\end{equation}
We expect waves to somehow represent elementary particles; for this
to be possible they must be compatible with the velocity definition
(\ref{eq:momentvel}) and Dirac's equation. The latter is a first
order differential equation, consequently we must establish a first
order wave equation from which one can derive the second order one,
velocity and Dirac equation. One possible first order equation is
\begin{equation}
    \label{eq:1stwave}
    (\Box - i\partial_t) \psi = 0.
\end{equation}
This produces Eq.\ (\ref{eq:4dwave}) if we multiply on the left by
$(\Box - i\partial_t)$. Equation (\ref{eq:1stwave}) is also the
condition for $\psi$ to be a monogenic function \cite{Doran03}.

The wave equation allows for harmonic solutions and consequently we
try the following solution
\begin{equation}
    \label{eq:psidef}
    \psi = \psi_0 \mathrm{e}^{u (\pm p_0 t + p_\mu x^\mu)};
\end{equation}
here $u$ is a square root of $-1$ whose characteristics we shall determine,
$p_0$ is the wave angular frequency and $p_\mu$ are components of a generalised
wave vector. When this solution is inserted in the 1st order equation
(\ref{eq:1stwave}) we get
\begin{equation}
    \label{eq:verifysol}
    (g^\mu p_\mu  \mp i p_0) \psi_0 u = 0.
\end{equation}
The first member can only be zero if $\psi_0$ is a multiple of the vector in
parenthesis and is nilpotent, i.e.
\begin{equation}
    \label{eq:nilpotent}
    g^{\mu\nu}p_\mu p_\nu -(p_0)^2 =0.
\end{equation}
We note here that Rowlands has been proposing a nilpotent formulation of Dirac
equation for some years, albeit with a different algebra \cite{Rowlands03}.

The velocity definition is incorporated in Eq.\ (\ref{eq:verifysol}) sufficing
for it to assign the conjugate momentum components to the wave vector by the
relation
\begin{equation}
    v_\mu = \frac{p_\mu}{p_0}.
\end{equation}
In order to recover Dirac's equation from Eq.\ (\ref{eq:1stwave}) we
have to consider the field free situation, which amounts to
replacing the curved vector derivative $\Box$ by its Euclidean
counterpart $\nabla$; the equation then becomes $(\nabla - i
\partial_t) \psi = 0$. Multiplying on the left by $\sigma_4$ we get
\begin{equation}
    (-\sigma_4 i \partial_t + \sigma_{4j} \partial_j + \partial_4) \psi=0.
\end{equation}
Now note that $\sigma_4 i$ squares to unity and $\sigma_{4j}$
squares to minus unity, so it is legitimate to make assignments to
the Dirac matrices: $\gamma^0 \equiv -\sigma_4 i$, $\gamma^j \equiv
\sigma_{4j}$. The last term in the previous equation must be
examined with consideration for the proposed solution; deriving
$\psi$ with respect to $x^4$ we get $\partial_4 \psi =  p_4 \psi_0 u
\exp[u(\pm p_0 t + p_\mu x^\mu)]$. Since $u$ will always commute
with the exponential, this simplifies to
\begin{equation}
    \partial_4 \psi = - p_4 \psi u.
\end{equation}
We can then make the further assignments $p_0 = E$, $p_4 = m$ and write
\begin{equation}
    \label{eq:Dirac}
    \gamma^{\bar{\mu}}\partial_{\bar{\mu}} \psi = - m \psi u.
\end{equation}

Dirac's equation has been written in a similar form by
\citet{Hestenes75, Hestenes86}, \citet{Doran93} and \citet{Doran03};
in all cases the authors chose $u = \sigma_{12}$ but we feel this to
be a rather restricting option. In fact there is no reason why one
should not be able to use any of the 12; 23; 13 choices for the
bivector index or even bivector combinations; for instance $u =
(\sigma_{12} + \sigma_{23} + \sigma_{13})/ \sqrt{3} $ seems a
perfectly reasonable choice, with the advantage that it is symmetric
in the 3 spatial coordinates. Pending further studies we propose the
following associations for the elementary particles:
\begin{itemize}
    \item down quarks: $u = \sigma_{32}$ and permutations;
    \item up quarks: $u = (\sigma_{23} + \sigma_{31})/\sqrt{2}$ and permutations;
    \item electron: $u =(\sigma_{21} + \sigma_{32} + \sigma_{13})/ \sqrt{3}
    $.
\end{itemize}
Electric charge is obviously encoded by $1/3$ the number of basis
bivectors intervening in $u$ with charge sign being given by the
direct or reverse order of the vectors. Consistently with the former
assignments we propose that anti-particles use the basis bivectors
symmetric to those listed above. For instance $u = \sigma_{32}$ is a
down quark with charge $-1/3$ and $u = \sigma_{23}$ is its
anti-particle with charge $+1/3$. Spin is associated with the $\pm$
sign in the exponent of $\psi$, one of the signs for up spin and the
other for down spin.
\section{Conclusion}
Our point of departure is the assumption that physics will one day
become indistinguishable from geometry, once unification and true
understanding of physics has been achieved. The exposition was
essentially geometric, demonstrating that the 4-dimensional space
that can be obtained from 5-dimensions with signature $(-++++)$ by
imposition of null displacement length, incorporates relations that
are the same as found in the main physical theories.

It was shown that general relativity metrics can be derived in such
space, as well as the metrics of 4D Euclidean space; a conversion
formula between the two spaces' metrics was derived for stationary
metric cases. The equations relating space curvature to its sources
were investigated for Euclidean space and solved for the case of
spherically symmetric mass. The solution that was found is PPN
equivalent to Schwarzschild's, although it will produce very
different predictions for large $M/r$.

The study of geodesics in 4D Euclidean space is equivalent to the
extension of geometric optics to 4 dimensions, justifying the
designation \emph{4-dimensional optics}. This extension would not be
complete is it did not apply to wave optics. It is shown that a 4D
wave equation can be obtained from the monogenic condition applied
in 5D. The latter is also shown to produce Dirac equation and to be
compatible with the dynamics equations established before.

A somewhat speculative encoding for the seven elementary particles
of the first generation is also proposed, based on the multiple
square roots of $-1$ present in the algebra. This is a subject to be
developed in forthcoming publications.
\end{multicols}

\centerline{\rule{7cm}{0.5pt}}
%
  \bibliographystyle{unsrtbda}   
  \bibliography{Abrev,aberrations,assistentes}   
%
\end{document}